\documentclass{Interspeech2024}

\usepackage[justification=centering,font=scriptsize,skip=0pt]{subcaption}
\graphicspath{{./Figures/}}
\usepackage{comment}
\usepackage{graphicx}
\usepackage{multirow}
\usepackage{cite}
\usepackage{color, colortbl}
\definecolor{Gray}{gray}{0.9}
\interspeechcameraready

\title{A Layer-Anchoring Strategy for Enhancing Cross-Lingual Speech Emotion Recognition}
\name[affiliation={1}]{Shreya G.}{Upadhyay}
\name[affiliation={2}]{Carlos}{Busso}
\name[affiliation={1}]{Chi-Chun}{Lee}

\address{
  $^1$National Tsing Hua University, Taiwan\\
  $^2$University of Texas at Dallas, USA}
\email{shreya@gapp.nthu.edu.tw, busso@utdallas.edu, cclee@ee.nthu.edu.tw}


\begin{document}

\maketitle
 
\begin{abstract}
Cross-lingual \emph{speech emotion recognition} (SER) is important for a wide range of everyday applications. While recent SER research relies heavily on large pretrained models for emotion training, existing studies often concentrate solely on the final transformer layer of these models. However, given the task-specific nature and hierarchical architecture of these models, each transformer layer encapsulates different levels of information. Leveraging this hierarchical structure, our study focuses on the information embedded across different layers. Through an examination of layer feature similarity across different languages, we propose a novel strategy called a layer-anchoring mechanism to facilitate emotion transfer in cross-lingual SER tasks. Our approach is evaluated using two distinct language affective corpora (MSP-Podcast and BIIC-Podcast), achieving a best UAR performance of 60.21\% on the BIIC-podcast corpus. The analysis uncovers interesting insights into the behavior of popular pretrained models. 

\end{abstract}

\section{Introduction} 

Recently, large pretrained models like Wav2Vec 2.0 \cite{10.5555/3295222.3295349}, WavLM \cite{chen2022wavlm}, Whisper \cite{radford2023robust}, and Hubert \cite{hsu2021hubert} have gained popularity, offering versatile capabilities across diverse applications. Trained on extensive datasets, these models serve as potent resources for tasks beyond their original domains. A notable trend involves fine-tuning these pretrained models for specific tasks such as SER \cite{upadhyay2023phonetic, chen2023exploring, pepino2021emotion, feng2023peft}, phonetics \cite{tom2022wav2vec, english2022domain}, speaker or language change detection \cite{berns2023speaker}, and speaker identification \cite{wang2021fine, fan2021exploring}. These models also serve as feature extractors, providing  a rich source of abstract representations useful across different tasks.

These transformer-based models exhibit a hierarchical architecture and within this hierarchy, diverse levels of information are embedded in the transformer layers \cite{li2023exploration}. The layer information varies based on the task specificity. For example, in models trained for \emph{automatic speech recognition} (ASR), initial layers capture fundamental acoustic details, while later layers encapsulate more complex lexical information. When employing these layer embeddings for different tasks, the ability to selectively choose or assign weights to layers that are more relevant can enhance the learning. Numerous studies not only use the final transformer layer but also utilize the information within other layers of the pretrained models \cite{peng2023attention , popov2018practices, english2022domain}. They aim to utilize this valuable information in a weighted or averaged manner, aligning features more effectively with specific task requirements. For instance, in speaker verification tasks \cite{peng2023attention}, various studies analyze different feature extraction methodologies developed upon pretrained models. Additionally, they explore regularization techniques and learning rate scheduling to stabilize the fine-tuning process, resulting in notable performance enhancements. Another research endeavor \cite{popov2018practices} emphasizes that speech representations derived from specific neural models like Transformers exhibit closer alignment with human perception, particularly regarding phonetic transcriptions. Moreover, English et al. \cite{english2022domain} highlights the Transformer architectures' capacity to effectively capture significant phonetic nuances.

\begin{figure*}[!ht]    
\centering
    \begin{subfigure}{.6\textwidth}
       \includegraphics[width=\textwidth]{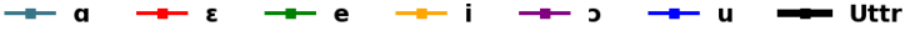}
    \end{subfigure}%
    
    \begin{subfigure}{.24\textwidth}
       \includegraphics[width=\textwidth]{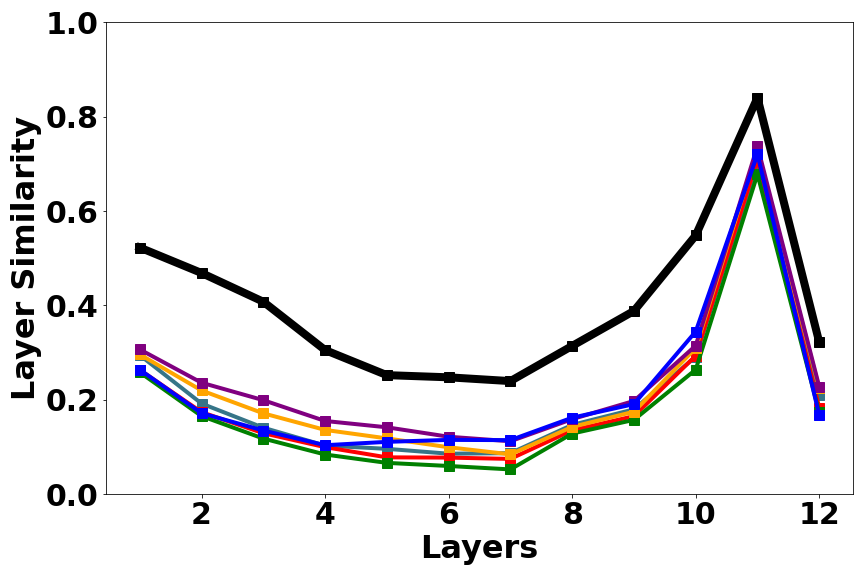}
       \subcaption{\emph{WavLM-Neutral}}
       \label{fig:wav-neu}
    \end{subfigure}%
    \begin{subfigure}{.24\textwidth}
        \includegraphics[width=\textwidth]{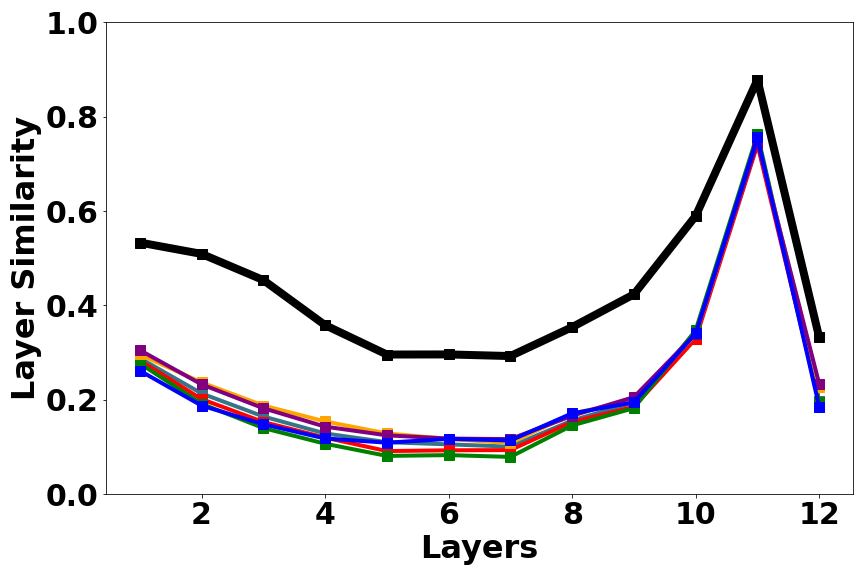}
        \subcaption{\emph{WavLM-Happiness}}
        \label{fig:wav-hap}
    \end{subfigure}%
    \begin{subfigure}{.24\textwidth}
       \includegraphics[width=\textwidth]{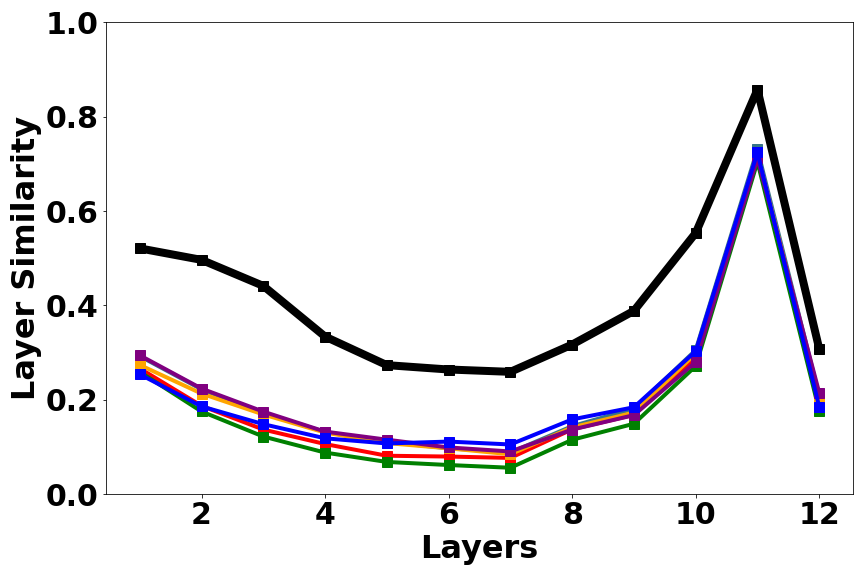}
       \subcaption{\emph{WavLM-Anger}}
       \label{fig:wav-ang}
    \end{subfigure}
    \begin{subfigure}{.24\textwidth}
        \includegraphics[width=\textwidth]{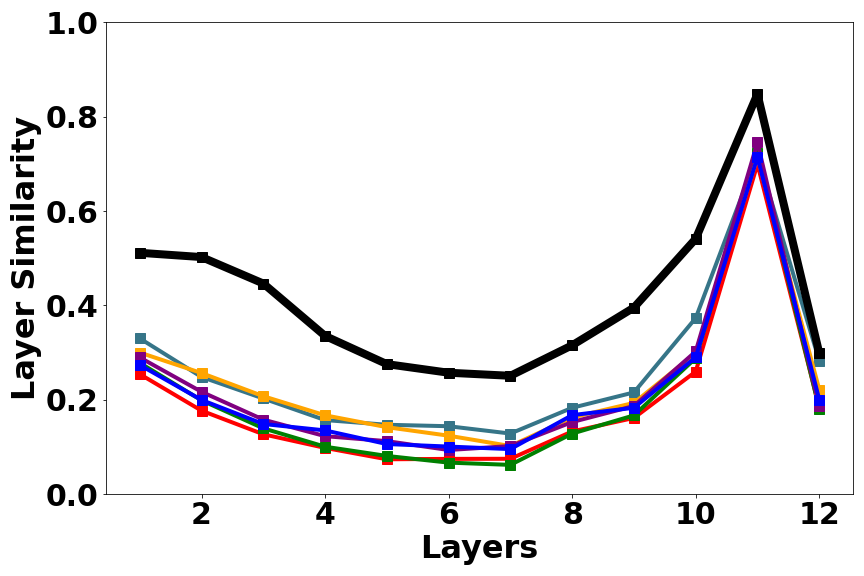}
        \subcaption{\emph{WavLM-Sadness}}
        \label{fig:wav-sad}
    \end{subfigure}%

    \begin{subfigure}{.24\textwidth}
       \includegraphics[width=\textwidth]{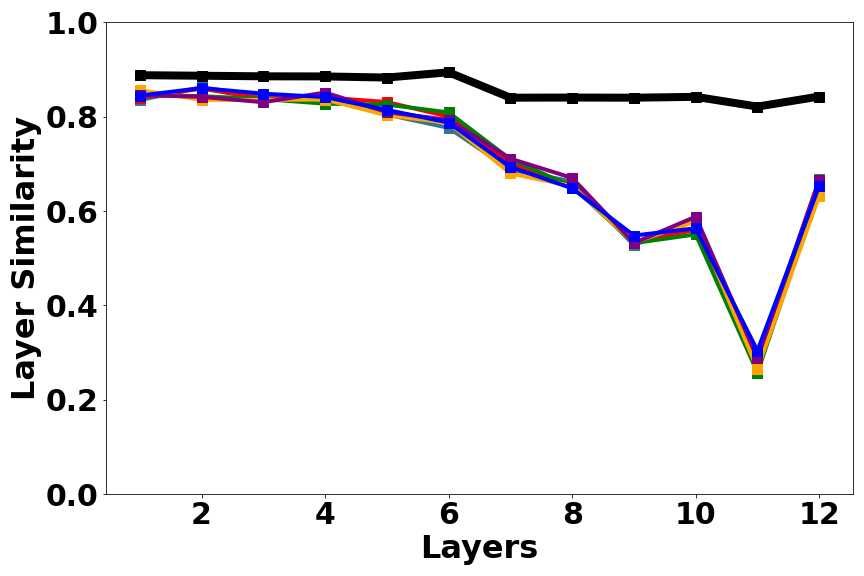}
       \subcaption{\emph{Whisper-Neutral}}
        \label{fig:whis-neu}
    \end{subfigure}%
    \begin{subfigure}{.24\textwidth}
        \includegraphics[width=\textwidth]{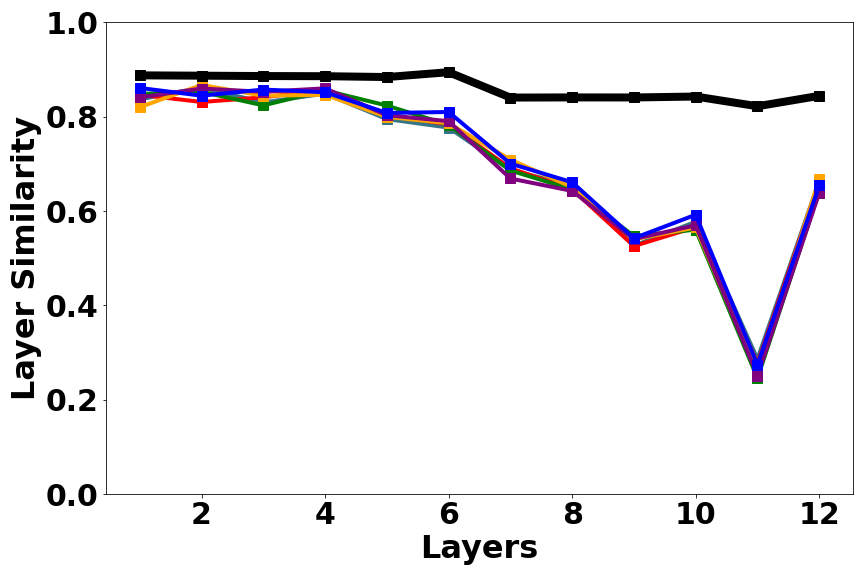}
        \subcaption{\emph{Whisper-Happiness}}
        \label{fig:whis-hap}
    \end{subfigure}%
    \begin{subfigure}{.24\textwidth}
       \includegraphics[width=\textwidth]{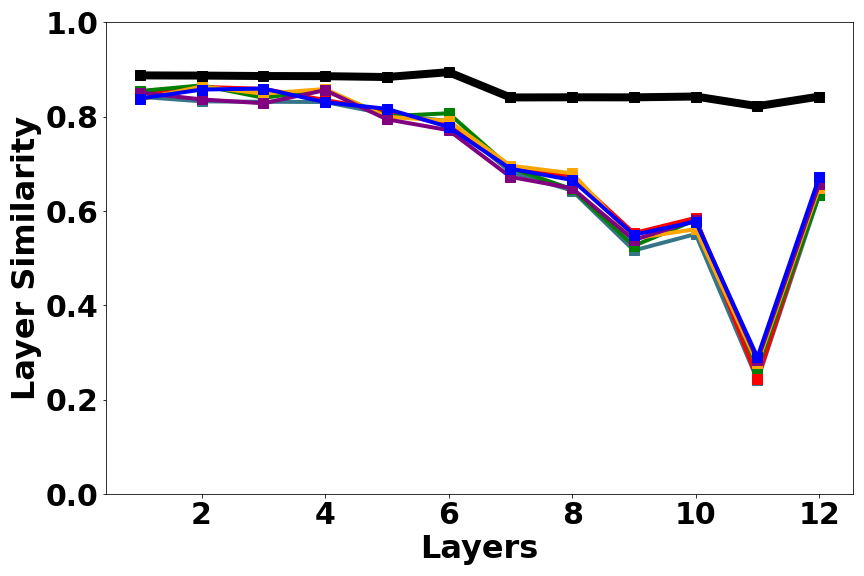}
       \subcaption{\emph{Whisper-Anger}}
       \label{fig:whis-ang}
    \end{subfigure}
    \begin{subfigure}{.24\textwidth}
        \includegraphics[width=\textwidth]{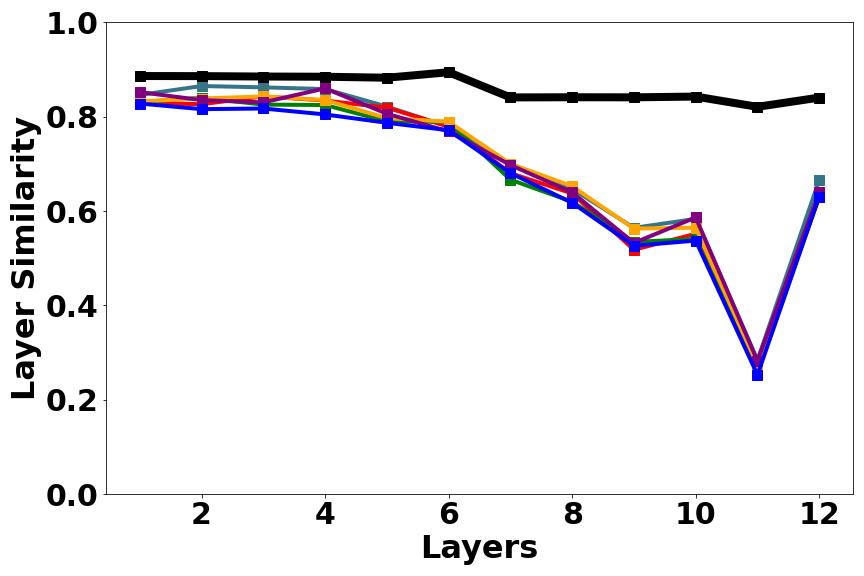}
        \subcaption{\emph{Whisper-Sadness}}
        \label{fig:whis-sad}
    \end{subfigure}%
\caption{Phonetic similarities across feature layer representations in MSP-P and BIIC-P corpora for WavLM and Whisper models, illustrating vowel-level and utterance-level phonetic similarities.}
\label{fig:feature_space}
\vspace{-0.4cm}
\end{figure*}

Speech representations derived from specific neural models, such as different transformer layers in pretrained models exhibit enhanced efficiency in recognizing phoneme \cite{bartelds2022neural}. These pretrained model's transformer architectures are adept at capturing substantial levels of phonetic information across various layers \cite{bartelds2022neural, english2022domain}. In the domain of cross-lingual SER, where phonetic alignment between diverse language corpora is advantageous \cite{upadhyay2023phonetic}, certain layers within large pretrained models beyond the final layer may hold greater importance \cite{li2023exploration}. These layers which directly encode acoustic cues and phonetic characteristics can form the fundamentals of tasks related to emotion recognition. Recognizing the potential efficiency gains in cross-lingual SER models, this paper introduces a novel approach known as \emph{Layer-anchoring}. This method strategically aligns layers based on their similarity across the two language corpora. By prioritizing and aligning layers that exhibit greater commonality between the features of both corpora, the model can enhance its performance. This novel strategy acknowledges the task-specific nature of SER and leverages the hierarchical structure of pretrained models to optimize layer representation utilization in a nuanced and contextually relevant manner.

Given that WavLM \cite{chen2022wavlm} has currently secured the top position on the SUPERB benchmark \cite{ shi2023ml} (retrieved on March 10, 2024), our experimentation will focus on utilizing WavLM and analyzing its performance with our proposed layer-anchoring algorithm. To address cross-lingual scenarios, we also use the multilingual pretrained model, Whisper \cite{radford2023robust} a widely adopted in recent studies \cite{vasquez2023novel, zezario2023study}, to compare its performance and insights with those of the monolingual model (WavLM). This study employs two distinct language corpora: the MSP-Podcast (American English) \cite{lotfian2017building} and the BIIC-Podcast (Taiwanese Mandarin) \cite{upadhyay2023an} corpora. First, we analyze layer similarities within the pretrained model's encoded features across both corpora. This analysis reveals layers exhibiting better commonality between the corpora than the final layer. Building upon this insight, we implement the \emph{layer anchoring mechanism} (LAM) to develop a cross-lingual SER model. Our proposed model, referred to as a layer-anchoring mechanism with a group of layers (LAM-GL), outperforms alternative approaches achieving 60.21\% \emph{unweighted average recall} (UAR) with WavLM features and 59.65\% with Whisper encoded features over the BIIC-podcast corpus.

\vspace{-0.05cm}
\section{Layer Similarity Analysis}
\label{subsec:phonetic_analysis}

\subsection{Naturalistic Corpora}
\vspace{-0.1cm}
\textbf{The MSP-Podcast (MSP-P)} \cite{lotfian2017building} corpus contains 166 hours of emotional \emph{American English} speech (v1.10), sourced from audio-sharing websites. This resource is valuable for SER research due to its extensive size and emotionally balanced dialogues from various individuals. It includes annotations for primary emotions, secondary emotions, and emotional attributes.  
In this study, this corpus is used primarily as the source corpus and focuses only on four primary emotion categories (\emph{Neutral, Happiness, Anger}, and \emph{Sadness}), comprising a total of 49,018 samples with predefined train-validation-test splits. The phonetic information is already included in the MSP-P corpus.

\smallskip
\noindent
\textbf{The BIIC-Podcast (BIIC-P)} \cite{upadhyay2023an} corpus is a SER database (v1.0) in \emph{Taiwanese Mandarin}. It contains 157 hours of speech samples from podcasts and follows a data collection methodology similar to the MSP-P corpus. The annotations cover primary and secondary emotional categories, as well as three emotional attributes. Here, the BIIC-P corpus is used as the target corpus. For this study, we employ 22,799 samples focusing on four primary emotion categories with the predefined train-validation-test splits given by the database provider. For the BIIC-P corpus phonetic knowledge, we employ the same phone aligner as shown in our previous work \cite{upadhyay2023phonetic}.

\vspace{-0.1cm}
\subsection{Layer Similarities}
\label{subsec:phonetic_sim}
\vspace{-0.1cm}
Previous research using large pretrained models on different tasks reveals that each layer contributes different levels of information during task learning \cite{bartelds2022neural, english2022domain}. Building on these findings, we aim to examine the degree of layer-similarity encoded within these model's layer representations across two distinct language corpora (the MSP-P and the BIIC-P). To explore this comparison, we employ two off-the-shelf models: WavLM and Whisper. We analyze the layer-similarity at both utterance and phonetic levels. We include phonetic-level analysis because our prior study \cite{upadhyay2023phonetic} has shown that cross-lingual contexts may reveal shared phonetic commonalities. Figure \ref{fig:feature_space} visually illustrates the cosine similarity between layer representations of the BIIC-P and MSP-P corpora across four primary emotions, including both the utterance-level and phonetic-level. Here, we only use training samples, excluding the test samples.

\smallskip
\noindent
\textbf{Utterance-Level Layer Similarity:} 
To assess the presence of similarities across layers in the two corpora for different emotions over the whole utterance, we extract all-layer feature representations from the considered pretrained models (WavLM and Whisper) for both the MSP-P and the BIIC-P corpora using entire utterances. Subsequently, we compute the layer-similarity using the \emph{cosine similarity} metric between the layer representations of the MSP-P and the BIIC-P corpora across each emotional category as presented in Figure \ref{fig:feature_space}. 
Upon examining plots depicted in Figure \ref{fig:feature_space}, we observe differing layer similarity behaviors between the WavLM and Whisper models' feature embeddings. In WavLM, later layers, such as layer 11 show higher similarity, while the Whisper model's initial layers (1 to 5) exhibit greater similarity. This trend is constant across different emotions. WavLM shares a similar phenomenon with wav2vec2.0 \cite{li2023exploration}, suggesting that high-level features may be more general and potentially lead to higher similarity. However, this assumption contrasts with the Whisper model, possibly due to distinct training methodologies. These findings prompt further exploration into phonetic-level layer similarity to determine any distinct observations compared to utterance-level analysis.

\smallskip
\noindent
\textbf{Phonetic-Level Layer Similarity:}
For phonetic-level layer similarity analysis, our specific focus lies on vowels which according to literature possess a higher capacity to convey emotion and are prevalent across different languages. We consider six common vowels across the MSP-P and BIIC-P corpora: [\textipa{A}/\textipa{a}, \textipa{E}, \textipa{@}, \textipa{i}, \textipa{O}, \textipa{u}]. From Figure \ref{fig:feature_space}, a clear pattern emerges in the similarity of layer features at the vowel level between the WavLM and Whisper models, aligning with the observations made at the utterance-level. Specifically, the WavLM model displays a trend of increasing layer similarity in its later layers for all vowel phones, contrasting starkly with the Whisper model, where greater similarity is observed in the initial layers. Investigating inter-vowel segment similarities within [\textipa{A}/\textipa{a}, \textipa{E}, \textipa{@}, \textipa{i}, \textipa{O}, \textipa{u}], Figure \ref{fig:feature_space} reveals variations across layers for different emotions. Nonetheless, an overall high degree of similarity is observed between the corpora at the corresponding layers.

\smallskip
\noindent
The above analyses highlight that due to task specificity and the hierarchical nature of models, in self-supervised learning models like WavLM, later layers encapsulate more abstract patterns and language-specific phonetic nuances as the model learns to predict future speech tokens. Conversely, Whisper being weakly supervised, the early layers may capture more basic acoustic features as they primarily rely on the input data with explicit labels so we observe greater layer-similarity towards the initial layers. This observation indicates that similar layer selection relies not only on task specifics but also on the model's training methodology. The dissimilarity of the final layer could stem from its alignment with the pre-training objective, which prioritizes tasks other than SER. Thus, while effective for its original purposes, it might not optimize cross-language speech emotion recognition (CL-SER).

\vspace{-0.1cm}
\subsection{Unified Layer Selection}
\vspace{-0.1cm}
As per our hypothesis, aligning layer features exhibiting high layer-similarities across different language corpora and imposing constraints on those layers can enhance the effectiveness of emotion transfer in CL-SER tasks. To anchor on more similar layers, we select specific sets of layers based on the findings of our previous analysis in Section \ref{subsec:phonetic_sim}. Table \ref{tab:selected_layers} outlines the selected layers under different settings. Drawing from our previous experience, we have observed that a group of anchors tends to outperform individual ones. Therefore, the table includes the \emph{group-layers} (GL), representing clusters of highly similar layers. Additionally, we identify the \emph{best-layer} (BL) and \emph{worst-layers} (WL), along with three sets of \emph{random-layers} (RL1, RL2, RL3). Except for BL, we select the top three layers for all cases concerning training CL-SER in the subsequent section.

\begin{table}[t]
\centering
\caption{Selected layer for WavLM and Whisper model.}
\renewcommand{\arraystretch}{1}
\resizebox{0.7\columnwidth}{!}{%
\begin{tabular}{c|c|c}
\toprule\specialrule{\cmidrulewidth}{0pt}{0pt}
 & WavLM & Whisper \\ \hline \hline
Group-Layers (GL) & [8, 9, 11] & [1, 2, 3] \\
Best-Layer (BL) & [11] & [2] \\
Worst-Layers (WL) & [5, 6, 7] & [7, 10, 11]\\ \hline
Random-Layers (RL1) & [2, 6, 9]   & [2, 6, 9]  \\
Random-Layers (RL2) & [1, 5, 12]  & [1, 5, 12] \\ 
Random-Layers (RL3) & [3, 7, 11] & [3, 7, 11]\\
\specialrule{\cmidrulewidth}{0pt}{0pt}\bottomrule                          
\end{tabular}}
\label{tab:selected_layers}
\vspace{-0.5cm}
\end{table}

\vspace{-0.05cm}
\section{Layer Anchored Cross-Lingual SER}
In all our experiments, we consider the MSP-P (source) and BIIC-P (target) corpora as benchmarks to test our idea. The WavLM and Whisper embedding feature vectors are used as the pretrained layer representations. For the CL-SER architecture,  we use the transformer with 4-fully connected layer architecture, Following the same model presented in our previous work \cite{upadhyay2023phonetic} with an attention-weighted layer feature pooling concept as our backbone SER architecture. We employ the Adam optimizer with a learning rate of 0.0001 and a decay factor of 0.001, and back-propagation is done with the cross-entropy loss function. The network undergoes a maximum of 70 epochs and a batch size of 64 with early stopping. To evaluate model performances, we use the UAR metric. Since both utterance-level and phonetic-level similarities yield similar insights from Section~\ref{subsec:phonetic_analysis}, we integrate LAM over the entire utterance.

\begin{figure}[tbp]
  \centering
  \includegraphics[height=0.6\linewidth,width=1.0\linewidth]{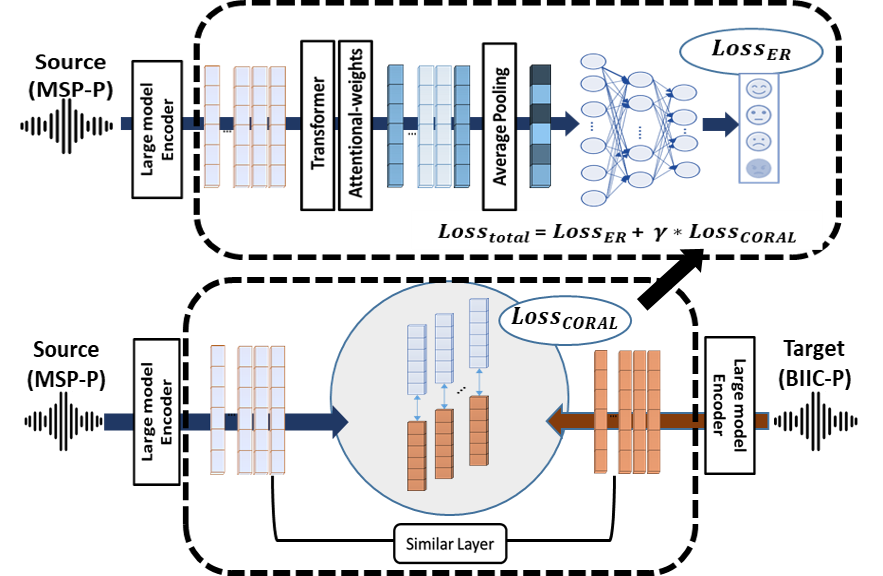}
  \caption{Proposed contrastive learning approach using layer anchoring mechanism (LAM) for cross-lingual SER.}
  \label{fig:arch}
  \vspace{-0.4cm}
\end{figure}

Our investigations detailed in Section~\ref{subsec:phonetic_analysis} offer initial insights suggesting that specific layers may exhibit more similarity and can enhance emotion modulation across both corpora. Motivated by these findings, we devise a layer anchoring mechanism aim at incorporating the layer-alignment constraint in the CL-SER modeling (Figure \ref{fig:arch}). Our proposed unsupervised CL-SER comprises two branches: (1) a conventional emotion classification branch tasked with classifying emotions, and (2) a layer anchoring mechanism (LAM) branch that identifies the layers in the transformer that increase the similarities across languages at the phonetic level. Equation~\ref{eq1} outlines the LAM loss.
\vspace{-0.3cm}
\begin{equation}
\label{eq1}
  {Loss}_{CORAL} = \sum_{i}^{N}{\lVert Cov ({L^{(i)}_{src}}) - Cov({L^{(i)}_{tar}}) ) \rVert}_F
\end{equation}
\vspace{-0.2cm}

Where \({L^{(i)}_{src}}\) and \({L^{(i)}_{tar}}\) denote the feature representations of layer \(i\) from the source and target corpora, respectively. \({Cov(\cdot)}\) represents the covariance matrix and \({\lVert . \rVert}_F \) denotes the Frobenius norm. Let \(n\) be the number of predefined layers for anchoring. The \({Loss}_{CORAL}\) is the Correlation Alignment Loss (CORAL) \cite{sun2017correlation} between the source and target representations for these layers. 

More specifically, in the LAM scenario where the CORAL loss is used, the source and target features refers to the similar layer representations from the MSP-P corpus and the BIIC-P corpus, respectively. This is to align the distributions of features between these two layer-similar representations by minimizing the difference in their second-order statistics.
The mathematical formulation for the attentional weighted average estimation is defined by Equation ~\ref{eq2},
\vspace{-0.3cm}
\begin{equation}
\label{eq2}
    {L}_{avg}=\sum_{i}^{12}{\alpha_{i}.{L^{(i)}_{src}}}
\end{equation}
\vspace{-0.2cm}

\noindent
where \({L}_{avg}\) denotes the attentional weighted average of the feature representations from all layers, and \(\alpha_{i}\) denotes the attention weight assigned to the layer \(i\). Here \(\sum_{i}^{12}{\alpha_{i}}=1\) and \(\alpha_{i} \geq 0\) for all \(i\). The weight \(\alpha_{i} \) can be computed using Equation \ref{eq4}.
\vspace{-0.2cm}
\begin{equation}
\label{eq4}
    \alpha_{i}={\frac{e^{a_i}}{\sum_{i}^{12}e^{a_j}}}
\end{equation}
\vspace{-0.2cm}

Where \(a_i\) represents the attention score for layer \(i\). The attention scores are the learnable parameter.

The overall loss for the first branch is the sum of the cross-entropy loss for the classic SER task and the CORAL loss for the layer anchoring mechanism. The complete loss is calculated using Equation~\ref{eq3},
\vspace{-0.2cm}
\begin{equation}
\label{eq3}
  L_{total}  = {Loss}_{ER} +  \gamma * {Loss}_{CORAL} 
\end{equation}
where \({L}_{ER}\) and \({L}_{CORAL}\) are the losses for the emotion classification and domain adaptation tasks. \(\gamma\) is the regularization parameter which is set to a constant value (\(\gamma\) = 0.5).

\vspace{-0.05cm}
\section{Experiment Results and Analyses}
\vspace{-0.05cm}
\subsection{Performance Comparison}
\vspace{-0.1cm}

Table \ref{tab:performance} shows the performance table, which includes results of our proposed idea LAM-GL, the baseline models, and the ablations results with WavLM and Whisper model's encoded layer representations. 

We assess three baseline methods: ensemble learning \cite{zehra2021cross}, which combines predictions from diverse models to enhance recognition accuracy in cross-lingual scenarios (Ensemble); few-shot learning \cite{ahn2021cross}, adapting models to target domains with limited labeled data (Few-shot); and our previously proposed Phonetic-constraint based anchoring (PA) method \cite{upadhyay2023phonetic}, used for learning in a common phonetic space for SER.
Compared with the baselines in Table \ref{tab:performance} for the MSP-P→BIIC-P task, our layer anchoring approach (LAM-GL) yields superior performance. Specifically, compared to models using only the last layer (the 12th layer), LAM-GL achieves a UAR of 60.21\%, surpassing Ensemble \cite{zehra2021cross} at 52.18\%, Few-shot \cite{ahn2021cross} at 53.62\%, and PC \cite{upadhyay2023phonetic} at 58.14\% with WavLM features. This enhanced performance is also evident with Whisper features. Additionally, drawing on prior work employing layer information for tasks like speaker identification and phone recognition \cite{peng2023attention}, we integrate these methods with PC, which outperforms Ensemble and Few-shot, denoted as PC-Avg \cite{upadhyay2023phonetic, peng2023attention} and PC-Atn \cite{upadhyay2023phonetic, peng2023attention}. Comparing these models with LAM-GL reveals further performance improvements, with WavLM features achieving 2.15\% and 1.38\%  UAR for PC-Avg and PC-Atn models, respectively. The same patterns are observed with Whisper features, suggesting that our LAM-GL model, which aligns more similar layers across various corpus features, provides improved utility for CL-SER.

\begin{table}[t!]
\centering
\caption{The proposed model performance (in UAR) for each SER task with considered baselines. It includes the statistical test over the baselines and the proposed LAM-GL model performances, denoted by asterisks (* for p \(<\) 0.1, ** for p \(<\) 0.05.}
\renewcommand{\arraystretch}{1.1}
\resizebox{0.45\textwidth}{!}{%
\begin{tabular}{c|c|cc|cc}
\toprule\specialrule{\cmidrulewidth}{0pt}{0pt}
                 & & \multicolumn{2}{c|}{MSP-P → BIIC-P} & \multicolumn{2}{c}{BIIC-P → MSP-P} \\ \hline
                 & & WavLM     & Whisper    & WavLM     & Whisper    \\ \hline \hline
                 \multirow{4}{*}{\rotatebox{90}{Top Layer}}
& CC  & 52.01\small{**}    & 51.87 \small{**}      &   48.39  \small{**}      &    49.01 \small{**}       \\
& Ensemble \cite{zehra2021cross}  &  52.18\small{**}     & 52.03\small{**}           &   51.75   \small{**}     &51.98  \small{**}          \\
& Few-shot \cite{ahn2021cross} &  53.62\small{**}      &  52.74\small{**}          & 50.59  \small{**}        &    51.75  \small{**}      \\ 
&PC \cite{upadhyay2023phonetic} & 58.14\small{**}    & 57.83\small{*}     &  55.35  \small{*}       & 54.93  \small{*}         \\ \hline
\multirow{10}{*}{\rotatebox{90}{w/ Layer}}
&PC-Avg \cite{upadhyay2023phonetic, peng2023attention} & 58.06\small{**}    & 58.01\small{**}      & 55.24   \small{*}       &  54.32 \small{*}         \\
&PC-Atn \cite{upadhyay2023phonetic, peng2023attention} & 58.83\small{*}    & 58.92\small{*}     &  55.64  \small{*}       &   56.10 \small{*}     \\  \cline{2-6}
&\textbf{LAM-GL} &\textbf{60.21}      & \textbf{59.65}     &  \textbf{ 56.68}        &  \textbf{56.37 }         \\ \cline{2-6}
&LAM-AL               & 58.54         & 57.97            &    55.39       &     54.91       \\
&LAM-BL               & 59.16         & 58.11            &   55.75        &  54.21          \\
&LAM-WL               &  58.01         & 57.72          &     54.64      &    53.77       \\
&LAM-RL1               &  58.94         & 56.24           &    54.93       &    54.29       \\
&LAM-RL2               &  58.55         & 57.39          &    53.85       &   53.38        \\
&LAM-RL3               &  57.23       & 57.84         &      54.20     & 54.43          \\
 \specialrule{\cmidrulewidth}{0pt}{0pt}\bottomrule   
\end{tabular}}
\label{tab:performance}
\vspace{-0.5cm}
\end{table}

To validate our LAM-GA method, we extended our investigation beyond the selected layers, exploring whether aligning any random layer of the two corpora or all layers is acceptable or if precise selection is necessary. We train LAM-GA models with diverse configurations: utilizing all layers (LAM-AL), the best layer (LAM-BL), the worst layer (LAM-WL), and random selections (LAM-RL1, LAM-RL2, LAM-RL3). Table \ref{tab:selected_layers} presents the selected layers for these analyses. The MSP-P→BIIC-P task results from Table \ref{tab:performance} indicate that anchoring all layer models (LAM-AL) does not enhance performance over LAM-GL, yielding 58.54\% for WavLM and 57.97\% for Whisper UAR. Similarly, using the best layer (LAM-BL), which achieves the UAR of 59.16\% for WavLM and 58.11\% for Whisper, is not better than the LAM-GL strategy. Furthermore, the performance of the worst layer (LAM-WL) suggests that aligning dissimilar layers can adversely affect model performance, with WavLM features scoring 58.01\% and Whisper features 57.72\%. Despite generating three random sets for both WavLM and Whisper model features, results obtained for LAM-RL1, LAM-RL2, and LAM-RL3 did not outperform our LAM-GA model. For instance, using WavLM features, we achieved 58.94\%, 58.55\%, and 57.23\% UAR with LAM-RL1, LAM-RL2, and LAM-RL3, respectively. This underscores the significance of precise layer selection for the LAM.

As a validation of our concept, we incorporate cross-lingual SER assessments utilizing the BIIC-P corpus as the source and the MSP-P corpus as the target. The results are presented in Table \ref{tab:performance}. In the BIIC-P→MSP-P task, our proposed model LAM-GL demonstrates superior performance compared to other models listed in Table \ref{tab:performance}, achieving 56.68\% with WavLM features and 56.37\% with Whisper features. The overall analysis of Table \ref{tab:performance} for the BIIC-P→MSP-P task confirms a similar trend to the one observed in the MSP-P→BIIC-P task.

\vspace{-0.05cm}
\subsection{CL-SER With Different Layer Selection Strategy} 
\vspace{-0.1cm}
In this investigation, we explore the performance of LAM-GL compared to LAM across various phoneme groups (vowel-based (\emph{Vowl}), consonant-based (\emph{Cons})), as well as an utterance-based approach (\emph{Uttr}) over specific emotion detection task. We segment utterances based on different phoneme groups (\emph{Vowl}, \emph{Cons}), selecting layers to train our LAM-GA model. 
Results in Table \ref{tab:analysis} for the MSP-P→BIIC-P task across four primary emotions reveal that while the selected layers remain relatively consistent across the \emph{Uttr}, \emph{Vowl}, and \emph{Cons} strategies, notable performance differences emerge across different emotions. Particularly, the \emph{Vowl} approach demonstrates superior performance for emotions such as \emph{Happiness} and \emph{Anger}, achieving 74.21\% UAR and 75.55\% UAR, respectively, with WavLM features. A similar trend is observed with the Whisper model. This significant finding suggests that vowels exhibit a higher level of commonality over the two different language corpora features, potentially facilitating more efficient emotion transfer compared to considering the entire utterance.

\begin{table}[t!]
\centering
\caption{Specific emotion recognition (in \% UAR) with different layer selection strategies.}
\renewcommand{\arraystretch}{1.1}
\resizebox{0.4\textwidth}{!}{%
\begin{tabular}{c|c|c|cccc}
\toprule\specialrule{\cmidrulewidth}{0pt}{0pt}
 & Model & Layers & Neu & Hap & Ang & Sad \\ \hline \hline
\multirow{2}{*}{\rotatebox{90}{Uttr}} &WavLM & [8,9,11] &75.13  & 72.88  &74.33  &69.57  \\
  &Whisper & [1,2,3]  & 75.70  & 72.63 & 73.80 &69.81  \\ \hline
 \multirow{2}{*}{\rotatebox{90}{Vowl}}   &WavLM& [8,9,11] &75.98  &\textbf{74.21}  &\textbf{75.55}  & 69.92  \\
  &Whisper& [1,3,5]  &74.83  &\textbf{75.02}  & \textbf{74.19} & 70.46  \\ \hline
\multirow{2}{*}{\rotatebox{90}{Cons}}   &WavLM & [9,10,11] & 74.19  &73.73  &74.92  & 68.63 \\
  &Whisper& [3,5,6]  & 73.34 & 73.97 &73.30  &69.95  \\ 
 \specialrule{\cmidrulewidth}{0pt}{0pt}\bottomrule   
\end{tabular}}
\label{tab:analysis}
\vspace{-0.5cm}
\end{table}

\vspace{-0.1cm}
\section{Conclusion}
This study introduces a novel approach that aims at reducing layer feature disparities between different language corpora through a layer anchoring strategy. By capitalizing on pretrained models and aligning similar layer features from the source language to those of the target language, we illustrate the efficacy of our method in harmonizing phonetic characteristics while mitigating discrepancies. Our experimentation and evaluation reveal that our layer-anchoring strategy (\emph{LAM-GA}) achieves the best UAR of 60.21\% by effectively facilitating emotion transfer in cross-lingual SER. Additionally, we uncover intriguing insights indicating that the selection of layers is not uniform across all pretrained models but varies depending on the task and the model's training methodology. Our future work will delve deeper into the observed differences in specific emotion recognition using the \emph{LAM-GA} method under various strategies, even when the layer disparities are minimal. Also, we will explore enhanced algorithms to accommodate multiple languages in the SER training.

\newpage

\section{Acknowledgements}
This work was supported by the NSTC under Grants 110-2634-F-002-050 and 110-2221-E-007-067-MY3, and the NSF under Grant CNS-2016719.

\bibliographystyle{IEEEtran}
\bibliography{template}

\end{document}